# ADWISE: Adaptive Window-based Streaming Edge Partitioning for High-Speed Graph Processing

Christian Mayer, Ruben Mayer, Muhammad Adnan Tariq, Heiko Geppert,
Larissa Laich, Lukas Rieger, and Kurt Rothermel
Institute of Parallel and Distributed Systems
University of Stuttgart, Germany
firstname.lastname@ipvs.uni-stuttgart.de

*Abstract*—In recent years, the graph partitioning problem gained importance as a mandatory preprocessing step for distributed graph processing on very large graphs. Existing graph partitioning algorithms minimize partitioning latency by assigning individual graph edges to partitions in a streaming manner — at the cost of reduced partitioning quality. However, we argue that the mere minimization of partitioning latency is not the optimal design choice in terms of minimizing total graph analysis latency, i.e., the sum of partitioning and processing latency. Instead, for complex and long-running graph processing algorithms that run on very large graphs, it is beneficial to invest more time into graph partitioning to reach a higher partitioning quality — which drastically reduces graph processing latency. In this paper, we propose ADWISE, a novel *window-based* streaming partitioning algorithm that increases the partitioning quality by always choosing the best edge from a *set of edges* for assignment to a partition. In doing so, ADWISE controls the partitioning latency by adapting the window size dynamically at run-time. Our evaluations show that ADWISE can reach the *sweet spot* between graph partitioning latency and graph processing latency, reducing the total latency of partitioning plus processing by up to $23 - 47$ percent compared to the state-of-the-art.

*Keywords*-Graph partitioning; vertex-cut; edge partitioning; adaptive; window; streaming; distributed graph processing

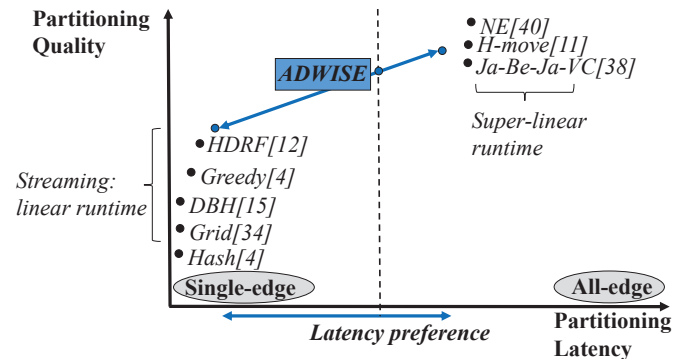

Fig. 1: Research gap – adaptive window-based streaming vertex-cut partitioning.

## I. Introduction

The last decade has brought a massive growth of graph-structured data. Web graphs link trillions of documents, social networks connect billions of users, recommendation graphs connect millions of people to billions of products, movies, or songs, and deep neural networks comprise of billions of highly connected artificial neurons. Analyzing the graphs *with low latency* is crucial for interactive recommendation queries using collaborative filtering [1] (*"Which movie to watch?"*), for online inference on graphical models using belief propagation [2] (*"How to rate this Go game position?"*), for PageRank to extract timely insights [3] (*"How does SEO impact our website's rank?"*), or simply to reduce costs of graph analysis in the cloud (e.g., AWS charges an hourly rate). To this end, specialized graph processing systems such as Pregel [3], PowerGraph [4], and GraphX [5] emerged that scale out computation by dividing the graph into multiple partitions to be processed in parallel by multiple worker machines.

In doing so, *vertex-cut* partitioning assigns each edge exclusively to a single worker machine, and thus, divides the graph along vertices. We focus on vertex-cut partitioning in this paper due to its superior partitioning properties on real-world graphs compared to edge-cut partitioning [4]. In vertex-cut partitioning, each vertex can reside on multiple partitions, i.e., can be replicated across the corresponding worker machines. However, a replicated vertex causes synchronization and communication overhead between the worker machines, inducing higher graph processing latency [2], [6], [7]. Hence, graph processing latency strongly correlates with *partitioning quality*, defined as the replication degree of vertices on the different worker machines.

The problem of partitioning a graph optimally, i.e., with minimal vertex replication, is impracticable for large graphs due to its NP-hardness [8]. In literature, there are two basic approaches to practically address the partitioning problem: (i) *single-edge* streaming algorithms perform partitioning decisions on one edge at a time, minimizing the partitioning latency, or (ii) *all-edge* algorithms load the complete graph into memory and employ global placement heuristics to optimize the partitioning quality. The existing algorithms follow either of the methods: Figure 1 illustrates the landscape of state-of-the-art vertex-cut partitioning algorithms. Modern graph processing systems use streaming partitioning when loading massive graphs due to their superior scalability and minimal runtime complexity [4], [9].

In this paper, we investigate whether it is always optimal to invest minimal partitioning latency as done by the established streaming partitioning algorithms. Clearly, there is a trade-off between partitioning latency and partitioning quality— and thus, graph processing latency. Our hypothesis is that for complex and long-running graph algorithms that run on large graphs, investing more than minimal time into graph partitioning leads to reduced total latency of graph partitioning *and* graph processing: To minimize the total latency, this trade-off must become controllable, i.e., the partitioning algorithm should be able to control the time invested into optimizing the partitioning quality. However, none of the current streaming partitioning algorithms allow for that.

To close this gap, we propose to consider a *window of edges* from the graph stream for making the partitioning decisions—instead of either a single edge or all edges. The basic idea is that considering more edges at a time enables improvements on the partitioning quality, but imposes a larger partitioning latency. While this is an intuitive idea, it poses a number of interesting research questions that need to be addressed: (1) How many edges should be taken into account when making a partitioning decision, i.e., how large should the window be? (2) Which of the edges should be assigned to which partition, i.e., how to design the *scoring* function that assigns the highest score to the best edge placement? (3) How to avoid unnecessary computations, i.e., how to limit score calculations to the high-potential edges in the window?

To address these questions, we developed ADWISE [10], a novel *window-based* streaming partitioning approach. Our main contributions are as follows.

- We employ methods to *automatically adapt* the window size at runtime in order to control the trade-off between partitioning latency and quality according to a partitioning latency preference.
- We propose a novel *scoring function* tailored to window-based partitioning. It considers multiple objectives – including diversity and skewness of the graph edges – to quantify partitioning decisions pertaining to the edges in the window.
- We employ a *lazy traversal* score calculation method that limits score (re-)calculations to a subset of most promising window edges in order to reduce partitioning latency on a given window.
- We introduce the *spotlight partitioning* optimization for parallel graph partitioning on multiple ADWISE instances. Spotlight partitioning reduces the spread of the partitioning instances such that each instance works on a disjoint set of partitions. This tremendously improves partitioning quality and can be applied on top of any existing streaming graph partitioning algorithm.
- Our evaluations show that for large-scale real-world graph processing problems, it is beneficial to invest more latency into partitioning in order to minimize the total latency. Using ADWISE, the total latency could be reduced by up to $23-47\%$ compared to traditional single-edge streaming partitioning algorithms.

We state the problem formulation in Section II, describe ADWISE in Section III, evaluate in Section IV, present related work in Section V, and conclude in Section VI.

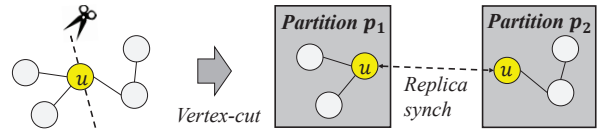

Fig. 2: Vertex-cut partitioning.

## II. PROBLEM STATEMENT AND ANALYSIS

In this section, we introduce the graph partitioning problem, define the window-based streaming partitioning model proposed by ADWISE, and discuss the research questions in window-based streaming partitioning that need to be solved.

### A. The Vertex-cut Graph Partitioning Problem

Many graph processing systems rely on vertex-cut partitioning [4], [5], [11], which we describe in the following. Let graph $G = (V, E)$ consist of a set of vertices $V = \{v_1, ..., v_n\}$ and edges $E \subseteq V \times V$. The goal is to divide the graph into $k$ partitions with identifiers $P = \{1, ..., k\}$. Vertex-cut graph partitioning can be achieved by *assigning edges to partitions*, which leads to cut vertices spanning multiple partitions. An example is given in Figure 2. The graph is cut through vertex $u$ into two partitions $p_1$ and $p_2$. Vertex $u$ is replicated on both partitions, because both contain edges incident to vertex $u$. We denote the set of partitions where vertex $u$ is replicated as *replica set* $R_u$. During graph processing, replicas of $u$ communicate to provide remote vertex data access to vertices residing on different partitions. By minimizing the number of replicas (denoted as *replication degree*), the amount of communication during graph computation is minimized as well [4]. Therefore, the goal of vertex-cut partitioning is to minimize the replication degree (cf. Equation 1), such that the partitions are balanced in the number of edges (cf. Equation 2) to ensure workload balancing during graph processing (cf. [12]). The maximal deviation between the number of edges assigned to any pair of partitions is controlled via the parameter $\tau \in [0, 1]$. In Table I, we give an overview about the notation used in this paper, in the order of occurrence.

$$minimize \frac{1}{|V|} \sum_{v \in V} |R_v|, \qquad (1)$$

$$s.t. \forall i, j \in P, |P_i| > |P_j| : \frac{|P_j|}{|P_i|} > \tau. \qquad (2)$$

### B. Streaming Partitioning

In the following, we analyze the streaming vertex-cut partitioning method in more detail, pointing out the commonalities and shortcomings of existing algorithms.

In vertex-cut streaming partitioning, partitioning algorithms perform a single pass over the stream of graph edges and assign all edges to partitions as they arrive in the stream. More precisely, given a sequence of edges $\langle e_1, ..., e_{|E|} : e_i \in E \rangle$, edge $e_i$ is assigned to partition $p_j \in P$ considering only previous assignment information from edges $\langle e_1, ..., e_{i-1} \rangle$. As each edge is accessed exactly once, the runtime complexity is linear to the number of edges.

| | |
|---|---|
| $G = (V, E)$ | Graph with set of vertices $V$ and edges $E$. |
| $P \subset \mathbb{N}$ | The set of partition ids. |
| $k \in \mathbb{N}$ | The number of partitions, i.e., $|P| = k$. |
| $R_u \subseteq P$ | Replica set of vertex $u$. |
| $P_i \subseteq E$ | The set of edges assigned to partition $i \in P$. |
| $\tau \in [0,1]$ | Maximal imbalance between any two partitions. |
| $g(e, p) \in \mathbb{R}$ | Score for edge $e$ and partition $p$. |
| $w \in \mathbb{N}$ | Number of edges in the window. |
| $W \subseteq E$ | The set of edges in the window with $|W| = w$. |
| $L \in \mathbb{N}$ | User-defined latency preference (milliseconds). |
| $S = \langle E \rangle$ | The edge stream, an ordered sequence of edges |
| $C \subseteq W$ | Set of high-score edges (*candidate set*). |
| $Q \subseteq W$ | Set of low-score edges (*secondary set*). |
| $\Theta \in \mathbb{R}$ | Score threshold to determine a candidate edge. |
| $B(p) \in \mathbb{R}$ | Balancing score of partition $p \in P$. |
| $\lambda \in \mathbb{R}$ | Balancing parameter and adaptive balancing function. |
| $R(e, p) \in \mathbb{R}$ | Replication score for $e \in W$ and $p \in P$. |
| $deg(v) \in \mathbb{N}$ | Degree of vertex $v$. |
| $N(u) \subseteq V$ | Set of neighbors of vertex $u \in V$. |
| $N_i(u) \subseteq V$ | Set of neighbors of vertex $u \in V$ on partition $i \in P$. |
| $CS(e, p) \in \mathbb{R}$ | Clustering score for $e \in W$ and $p \in P$. |

TABLE I: Notation overview.

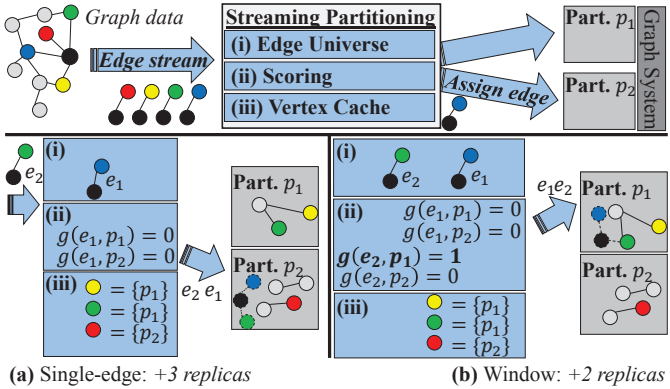

(a) Single-edge: *+3 replicas*  (b) Window: *+2 replicas*

Fig. 3: Streaming partitioning model.

We illustrate the streaming partitioning model at the top of Figure 3. The graph data is stored in a large file, a graph database, or a distributed file system. The streaming partitioning algorithm loads the data as a stream of graph edges and subsequently assigns them to partitions. Finally, these partitions are used for distributed graph processing. The streaming partitioning model consists of three building blocks. (i) The *edge universe* contains the set of edges from the stream that are considered for the partitioning decisions. Existing algorithms allow only one single edge in the edge universe. (ii) The *scoring function* measures how well an edge fits to a certain partition. (iii) The *vertex cache* maintains replica sets for all vertices that were assigned in any previous edge assignment. This information is used by the scoring function to determine the best edge assignment. All state-of-the-art streaming algorithms fit into this model, only the score computation differs.

**Shortcomings of single-edge streaming.** Due to the narrowness of the edge universe, existing partitioning algorithms enforce an assignment decision for each edge before populating the edge universe with the next edge. As a consequence, edge assignment decisions are often uninformed, i.e., based on insufficient knowledge about the replica sets of incident vertices. This can lead to low partitioning quality. Figure 3(a) provides an example. The scoring function $g(e_1, p_j)$ returns the number of times a vertex incident to edge $e_1$ is already replicated on partition $p_j$ (cf. [4]). Unfortunately, the vertex cache does not contain any information about the replica set of a vertex incident to edge $e_1$. Therefore, the score is zero for all partitions and the algorithm assigns edge $e_1$ to any partition (here: $p_2$) – an uninformed assignment decision. Next, the algorithm loads edge $e_2$ into the edge universe and assigns it to partition $p_2$ as selected by the scoring function. The assignment of both edges $e_1$ and $e_2$ leads to three new replicas (black, blue, and green vertex) on partition $p_2$.

### C. Window-based Streaming Partitioning

To overcome the uninformed assignment problem, the narrowness of the edge universe must be widened. When more edges are available in the edge universe, the partitioning algorithm can choose *which* edge to assign next to which partition. This is the basic idea of window-based streaming partitioning, as proposed by ADWISE. In the following, we extend the streaming partitioning model by the proposed windowing mechanism and point out the research questions that need to be solved.

*1) Basic Approach:* To improve partitioning quality, ADWISE extends the edge universe to contain multiple edges and iteratively assigns the edge with the highest score in the edge universe – thus *preferring informed and delaying uninformed edge assignments*. While the partitioning algorithm assigns more edges, it enriches the vertex cache with more information about the replica sets. Finally, the algorithm has gathered enough information for many of the previously uninformed edges. For example, in Figure 3(b), the edge universe contains edges $e_1$ and $e_2$. The scoring function prefers assignment of edge $e_2$ to partition $p_1$ because an incident vertex is already replicated on $p_1$ (green vertex). By assigning edge $e_2$ first (i.e., *before* $e_1$), the algorithm learns relevant information for edge $e_1$ ("black vertex replicated on $p_1$"). It assigns edge $e_1$ to partition $p_1$ and has saved one replica compared to the single-edge streaming algorithm.

*2) Research Questions:* Introducing a window to the streaming edge partitioning model will only improve the total latency when the window-based partitioning algorithm is carefully designed. This is a challenging task that has not been addressed in literature yet. In particular, the following questions have to be addressed.

*(1) How to set and adapt the optimal window size?* Although partitioning quality can be improved by increasing the window size $w \in \mathbb{N}, w \geq 1$, this also incurs more score computations leading to higher partitioning latency. There is a complex relation between partitioning latency, partitioning quality and graph processing latency. To be able to optimize the total latency, it is necessary that the partitioning latency can be controlled, i.e., a preference on partitioning latency can be set. How this can be achieved has not been investigated in previous works, as the single-edge streaming partitioning algorithms do not allow for such a degree of freedom.

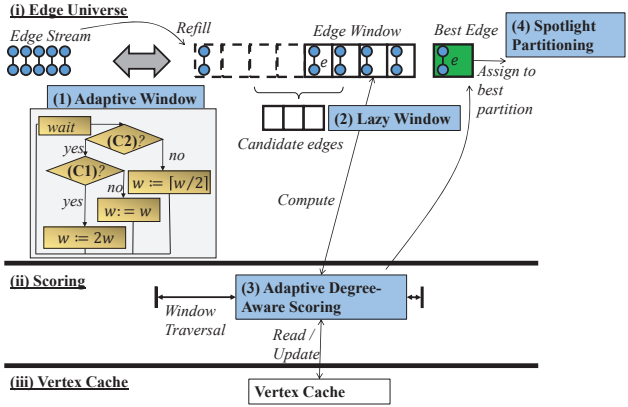

Fig. 4: Approach overview ADWISE.

*(2) How to reduce computational complexity of partitioning?* Calculating a score for each edge-partition pair in the window from scratch would lead to $\mathcal{O}(w)$ times the computational complexity of single-edge streaming algorithms. However, from-scratch calculations might not always be necessary because of significant computational overlap between two consecutive windows. An efficient window traversal algorithm should only compute the significant score deltas to the previous window.

*(3) How to tailor the scoring function to <u>window-based</u> streaming?* The scoring function in window-based streaming partitioning should effectively exploit the window's main advantage: the ability *to choose among multiple edges*. This increases flexibility – but only for carefully designed scoring functions that account for this additional dimension. Existing scoring functions from single-edge streaming partitioning can only decide about the best partition for a given edge.

In developing ADWISE, we have thoroughly investigated these research questions and have developed practical solutions, as described in the following section.

## III. ADWISE

ADWISE, the **AD**aptive **WI**ndow-based **S**treaming **E**dge partitioning algorithm, addresses the shortcomings of single-edge streaming algorithms by extending the edge universe with multiple edges, thus enabling more flexibility in the edge assignment decisions. Figure 4 provides an overview of the ADWISE algorithm. The edge universe consists of a window of $w$ edges. ADWISE iteratively selects the best edge from the edge window, assigns it to the best partition, and refills the window from the edge stream to contain $w$ edges again. In the following, we outline the general approach and highlight the main concepts of ADWISE.

**(1) Adaptive Windowing:** ADWISE allows to control the partitioning latency by automatically adapting the window size $w$ at runtime such that the algorithm keeps a partitioning latency preference $L \in \mathbb{N}$ (specified in milliseconds) with high probability. In the presence of sufficient partitioning time, the window size is increased to maximize partitioning quality; if the latency preference $L$ is likely to be violated, the window size is decreased. Section III-A provides a detailed description.

**(2) Lazy Window Traversal:** ADWISE exploits the property that high-score edges in one window are likely to remain high-score edges in the subsequent window. Hence, complete re-computation of the whole window after each edge assignment would lead to redundancies. We developed the optimization of *lazy window traversal* that exploits this property by calculating scores only for a subset of high-score edges in the window (*candidate set*). Non-candidate edges are updated only if significant changes in the vertex cache require re-computation of individual scores (cf. Section III-B).

**(3) Adaptive Degree-Aware Scoring Function:** To exploit the freedom to choose among multiple edges in a window when making the partitioning decisions, we introduce our scoring function $g(e, p)$ in Section III-C. It consists of three parts: (a) An adaptive load balancing score, (b) a degree-aware score, and (c) a clustering score.

(a) The partitioning decision of single-edge streaming approaches is significantly influenced by the objective of *balancing the number of edges* among partitions (cf. Equation 2) [12]. However, we argue that balancing partitions is not equally important in each phase of the stream. We introduce our optimization of adapting at runtime how much the balancing objective influences the partitioning decisions, based on the relative progress in the stream and the present imbalance of the partitions. (b) The degree-aware score quantifies how good edge $e \in W$ in edge window $W \subset E$ fits to partition $p$ by taking into account information about current replica sets from the vertex cache. (c) The clustering score prioritizes assignment of edges towards the local communities of the incident vertices – exploiting the cliquishness of real-world graphs.

**(4) Spotlight Partitioning:** When multiple instances of a streaming edge partitioning algorithm work on different chunks of the graph in parallel, it is of great importance to carefully consider how many partitions are filled by the different workers (i.e., the spread). To address this problem, we propose our optimization *"Spotlight"* that is reducing the spread of each partitioner such that partitioners can maintain locality by working on their own set of partitions. Details are provided in Section III-D.

### A. Adaptive Window Algorithm

In the following, we explain our method for trading partitioning latency versus quality. The basic idea is to increase the window size as long as this leads to better partitioning quality while the latency preference $L$ can be met. Otherwise, we decrease (or keep) the window size. To decide whether the latency preference can be met, ADWISE measures the average latency $lat_w$ of *assigning a single edge* (for current window size $w$). The algorithm starts by setting the window size to $w = 1$. After assigning $w$ edges and updating the average edge assignment latency $lat_w$, the algorithm either increases, keeps or decreases the window size (cf. the flow diagram in Figure 4). More precisely, the window size is set to $w \leftarrow 2w$, if the following two conditions (**C1**) and (**C2**) are met. (**C1**) The *last* increasing of the window size led to better edge assignment decisions (quantified by averaging the score $g(e, p)$ over $w$ edge assignments). (**C2**) The latency preference $L$ can be met – assuming stable average latency and a known number of edges in the stream (the graph size is usually known or can be determined efficiently using line count on the graph file). In more detail, (**C2**) is true, if the average latency $lat_w$ is smaller than the maximal latency per edge assignment, i.e., $lat_w < \frac{L'}{|E'|}$, where $|E'|$ is the number of edges left in the stream and $L'$ is the time until the latency preference would be exceeded. This ensures that there is only a small risk of not meeting the latency preference. If the average latency is too large to meet the latency preference $L$, i.e., ($\neg$**C2**), the algorithm decreases the window size to $w \leftarrow \lceil w/2 \rceil$. Note that if the latency preference $L$ is too tight (e.g. 0 seconds), the algorithm decreases $w$ until $w = 1$ leading to single-edge streaming partitioning.

---

**Algorithm 1** Window-based streaming vertex-cut algorithm.

1: $W \leftarrow \{\}$ ▷ Set of window edges
2: $S$ ▷ Edge stream
3: $c \leftarrow 0$ ▷ Assignment counter
4: **while** $S \neq \emptyset$ **do**
5:     **if** $|W| < w$ **then** $W \leftarrow W \cup \{S.next()\}$
6:     $(\hat{e}, \hat{p}) \leftarrow$ GETBESTASSIGNMENT()
7:     assign $\hat{e}$ to partition $\hat{p}$
8: **function** GETBESTASSIGNMENT()
9:     $(\hat{e}, \hat{p}) \leftarrow argmax_{(e,p) \in W \times P} g(e, p)$
10:     $W \leftarrow W \setminus \{\hat{e}\}$
11:     **if** $c \mod w = 0$ **then**
12:         **if** (**C1**) $\wedge$ (**C2**) **then**
13:             $w \leftarrow 2w$
14:             **while** $|W| < w$ **do** $W \leftarrow W \cup \{S.next()\}$
15:         **else if** $\neg$(**C2**) **then**
16:             $w \leftarrow \lceil w/2 \rceil$
17:     $c \leftarrow c + 1$
18:     **return** $(\hat{e}, \hat{p})$

---

We give an algorithmic description in Algorithm 1. There are three global variables: the edge window $W$ (initially empty), the edge stream $S$, and an assignment counter $c$ tracking the number of assigned edges since the last window change. In lines 4-7, the algorithm performs the main loop: reading an edge from the stream and adding it to the window, retrieving the best edge-partition pair $(\hat{e}, \hat{p})$ from the window, and assigning edge $\hat{e}$ to partition $\hat{p}$. The algorithm retrieves the edge-partition pair $(\hat{e}, \hat{p})$ with highest score $g(\hat{e}, \hat{p})$ by iterating over all edges in the window $e \in W$ and all partitions in $p \in P$ (line 9). This edge is assigned to partition $\hat{p}$ and removed from the window (line 10). After $w$ edge assignments, the algorithm performs the described adaptive window procedure (lines 11-17) using the two conditions (**C1**) and (**C2**).

### B. Lazy Window Traversal

Clearly, the algorithm presented in the last section requires $w \times |P|$ score computations for each edge assignment resulting in large overhead for large window sizes $w$. In the following, we develop the idea of reducing runtime complexity by traversing only the set of high-potential edges in the window (denoted as *candidate set C*). Conversely, the *secondary set* $Q$ contains the rest of the edges in the window. As the high-score edge is probably among the candidates, we focus on computing scores mainly for the candidates to decide which edge to assign next. If we select the candidate edges right, we will perform exactly the same assignment decisions while having a much lower runtime complexity (for $|C| << |Q|$).

But how to decide which edges to include into the candidate set? First, if we load a new edge into the window, we calculate the maximal score $\hat{g}$ for assigning edge $e$ to any of the partitions $p \in P$. If this score is higher than a certain threshold $\Theta$ (see below), we add $e$ to the candidate set $C$, otherwise we add edge $e$ to the secondary set $Q$. Second, if the candidate set is empty, we calculate scores for all edges in the secondary set and add all edges whose maximal score is larger than $\Theta$. Third, if assigning an edge leads to the creation of a new replica, the replica set of a vertex changes. In this case, edges in the secondary set that are incident to the vertex with changed replica set are reassessed whether they can be added to the candidate set. We dynamically adjust the threshold $\Theta$ to the average score $g_{avg}$ of window edges: $\Theta = g_{avg} + \epsilon$ for a small $\epsilon \in [0, 1]$ with the idea of including only edges in the candidate set that have better than average score.

### C. Scoring Window Edges

The scoring function quantifies how *good* edge $e$ fits to partition $p$. However, existing single-edge scoring functions have two drawbacks. (1) They assume fixed parameter values that are chosen by domain experts. (2) They only address the problem of finding the best partition given an edge, but not the problem of finding the best edge in the window. Our scoring function extends the state-of-the-art by three optimizations to address these concerns: *adaptive balancing score*, *degree-aware window score*, and the *clustering score*.

**Adaptive Balancing:** The optimization constraint in Equation 2 requires balanced partitions. Therefore, single-edge scoring functions reinforce edge assignments towards partitions with less workload (i.e., number of edges) considering a balancing score $B(p)$ that measures the difference between partition $p$'s and the maximal workload (cf. Equation 3).

$$B(p) = \frac{maxsize - |p|}{maxsize - minsize + \epsilon}. \tag{3}$$

State-of-the-art single-edge partitioning approaches use a parameter $\lambda$ to regulate how much the balancing score influences the scoring function [12]. This parameter is defined by users or domain experts [12], [4], [13]. However, selecting this parameter is a challenging problem, because different graphs require different choices.

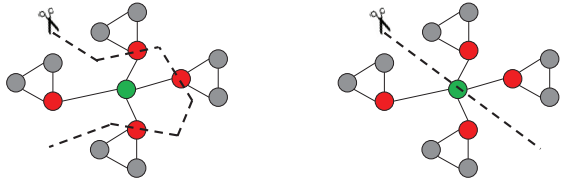

(a) Cut vertices with lower degrees  (b) Cut vertex with high degree

Fig. 5: Degree-aware vertex-cut partitioning.

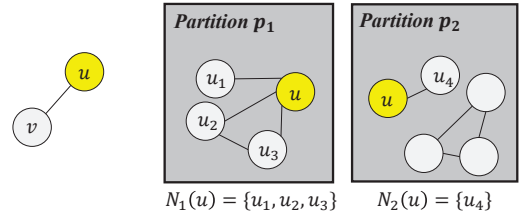

$N_1(u) = \{u_1, u_2, u_3\}$   $N_2(u) = \{u_4\}$

Fig. 6: Clustering Score Example.

To address this problem, we introduce an *adaptive balancing parameter* – releasing the user from the burden of choosing a suitable parameter in advance. We identified two requirements: (i) the balancing constraint can be relaxed in the beginning, i.e., $\lambda$ can be set to a small value, as long as there are still enough edges to compensate imbalanced partitions; (ii) if partitions are sufficiently balanced, a high parameter value for $\lambda$ distracts the scoring function from the main objective: minimize replication degree. Hence, our adaptive balancing parameter automatically adjusts to the current imbalance and progress of the partitioning algorithm. More precisely, we define the balancing parameter as a function $\lambda(\iota, \alpha)$ of the current imbalance $\iota = \frac{maxsize-minsize}{maxsize}$ and the fraction of already assigned edges $\alpha = min(1, \frac{|E'|}{m})$, where $E'$ is the set of already assigned edges and $m$ is the number of edges in the graph. Intuitively, the value of $\lambda(\iota, \alpha)$ should be low, i.e., tolerates high imbalance, if most edges are still unassigned. The highest acceptable imbalance, denoted as *tolerance*, should linearly decrease over time $\alpha$ as the end of the stream approaches, hence we define $tolerance(\alpha) = max(0, 1 - \alpha)$. If the current imbalance $\iota$ exceeds the tolerated imbalance (i.e., $\iota > tolerance(\alpha)$), balancing becomes more important and $\lambda(\iota, \alpha)$ should increase. Otherwise, balancing is currently not as important and $\lambda(\iota, \alpha)$ should decrease. In Equation 4, we specify our formula to set $\lambda(\iota, \alpha)$ adaptively after each edge assignment. To prevent extreme values, we keep $\lambda(\iota, \alpha)$ in the fixed interval $[0.4, 5]$.

$$\lambda_{new}(\iota, \alpha) = \lambda_{old}(\iota, \alpha) + (\iota - tolerance(\alpha)). \quad (4)$$

**Degree-aware Window Scoring:** The major objective is to minimize the replication degree (cf. Equation 1). Single-edge scoring functions use a replication score $R(e = (u,v), p)$ to quantify whether vertices $u$ and $v$ are already replicated on partition $p$ [12], [4], [13].

It is well-established that real-world graphs with skewed degree distributions can be divided well by preferably replicating high-degree vertices [14]. In Figure 5, we exemplify a stereotypical social network graph with high clustering coefficient for low-degree vertices and few high-degree vertices connecting the clusters. In Figure 5a, we cut the graph through vertices with median degree (red) leading to three replicated vertices. In Figure 5b, we cut the graph through the high-degree vertex (green) leading to only one replicated vertex.

Several approaches modify the replication score to consider *the relative vertex degree* of vertices $u$ and $v$ – in order to replicate high-degree before low-degree vertices [12], [15]. For instance, HDRF [12] maintains a degree table $deg$ with the current vertex degrees to calculate the *relative degree* of vertices $u$ and $v$, i.e., $\Psi'_u = \frac{deg(u)}{deg(u)+deg(v)} = 1 - \Psi'_v$. However, the *relative degree* of vertices incident to edge $e \in W$ lacks information about the *absolute degree* needed to differentiate window edges $e' \neq e \in W$. To resolve this, we introduce a truly degree-aware replication score by normalizing with respect to the vertex with maximal degree, i.e., $\Psi_u = \frac{deg(u)}{2maxDegree}$. With this modification, $\Psi$ returns low values for low-degree vertices *in the window*. To define the replication score, we use the *indicator function* $\mathbb{1}\{p \in R_u\}$ that returns 1 (or 0) if vertex $u$ is (or is not) replicated on partition $p$ (i.e., $p \in R_u$).

$$R((u,v),p) = \mathbb{1}\{p \in R_u\}(2-\Psi_u)+\mathbb{1}\{p \in R_v\}(2-\Psi_v). \quad (5)$$

**Clustering Score:** Many real-world graphs have a high local clustering coefficient (cf. *small-world networks*) [16]. Graph clustering algorithms that are able to identify the dense graph regions (i.e., *clusters*) can significantly increase locality and ultimately result in better partitioning quality [17].

How can we include this prior knowledge about strong local clusters into the scoring function? In Figure 6, we exemplify a simple scenario, where we have to decide whether edge $(u,v)$ should be assigned to partition $p_1$ or $p_2$. Vertex $u$ is already replicated on both partitions so the replication score does not help here and the partitions are balanced in the number of edges. However, vertex $u$ is already embedded into a strong local cluster on partition $p_1$, i.e., has three local neighbors $N_1(u) = \{u_1, u_2, u_3\}$, while it has only one local neighbor on partition $p_2$, i.e., $N_2(u) = \{u_4\}$. Intuitively, edge $(u,v)$ should be assigned to partition $p_1$ because edges $(v,x), x \in N_1(u)$ are likely to follow in the stream (*two friends of yours are more likely to be friends as well*).

In Equation 6, we define the clustering score $CS(e, p)$ for edge $e = (u,v)$ and partition $p$ as the number of times a neighboring vertex of $u$ or $v$ is already replicated on partition $p$, normalized to the interval $[0, 1]$. In the example, three neighbors of $(u,v)$, i.e., vertices $u_1, u_2, u_3$, are already replicated on partition $p_1$ compared to only one vertex $u_4$ on partition $p_2$ – leading to a higher clustering score for partition $p_1$. Note that for scalability reasons we calculate the neighboring function $N(u)$ for vertex $u$ only based on the vertices in the window, i.e., the larger the window, the more accurate is the clustering score.

| Name | $|V|$ | $|E|$ | $\hat{c}$ | Type |
|---|---|---|---|---|
| Orkut | 3,072,441 | 117,184,899 | 0.0413 | Social |
| Brain | 734,600 | 165,900,000 | 0.509766 | Biological |
| Web | 41,291,594 | 1,150,725,436 | 0.816026 | Web |

TABLE II: Real-world graphs for evaluations.

$$CS(e,p) = \frac{\sum_{u' \in N(u) \cup N(v)} \mathbb{1}\{p \in R_{u'}\}}{|N(u) \cup N(v)|} \quad (6)$$

Finally, we define the total scoring function of ADWISE in Equation 7.

$$g(e,p) = \lambda(\iota, \alpha) B(p) + R(e,p) + CS(e,p) \quad (7)$$

*D. Spotlight Partitioning*

To speedup partitioning, graph processing systems usually employ a parallel loading model, where each worker machine uses a separate, independent streaming graph partitioner [4], [7] – each processing a portion of the global graph (i.e., *chunk*) and filling its own vertex cache. Due to the limited information in each vertex cache, this leads to suboptimal partitioning decisions [4]. However, we identified a second reason for the worse replication degree which we denote as *spread of the partitioner*, i.e., the number of partitions each independent partitioner has to fill. If the spread is too large, the partitioner is forced to perform partitioning decisions mainly based on balancing considerations leading to increased replication degree. Roughly speaking, a large spread unnecessarily breaks up existing locality of edges in the edge stream.

Therefore, each of the $z$ partitioner divides its graph data among $\frac{k}{z}$ *exclusive* partitions. This simple optimization is extremely effective: it reduces replication degree by up to 80 percent (cf. Section IV) for all tested strategies while *reducing* computational overhead as well due to fewer score computations. Note that the resulting partitioning is still balanced (assuming equal-sized input chunks). Although this optimization seems straightforward, it has not been applied by previous partitioning algorithms.

## IV. EVALUATION

In this section, we evaluate different aspects of the ADWISE algorithm. First, we explore the trade-off between graph *partitioning* latency and *processing* latency. We show that ADWISE reduces the total graph latency (i.e., the sum of partitioning and processing latency) when computing standard graph processing algorithms on large real-world graphs. Then, we take a deeper look into parallel graph loading by analyzing the effects of the spotlight optimization on partitioning quality.

**Experimental Setup:** In our evaluations, we used three large real-world graphs Orkut [18], Brain [19], and Web [20] with up to 1.15 billion edges (cf. Table II). These graphs differ fundamentally with respect to the clustering coefficient $\hat{c}$: the social network Orkut has a rather weak clustering of $\hat{c} = 0.04$, the biological network Brain has moderate clustering of $\hat{c} = 0.51$, and Web has very strong clustering of $\hat{c} = 0.82$ (based on a graph sample [19]). We tested ADWISE on several smaller graphs [21] and a variable number of partitions and obtained similar (and in some cases even better) results. For brevity, we only state the results on large graphs with more than 100,000,000 edges.

As evaluation platform, we used an in-house computing cluster with 8 nodes × 8 Intel(R) Xeon(R) CPU cores (3.0GHz, 6144 KB cache) and 32GB RAM per node, connected via 1-Gigabit Ethernet. As benchmarks, we evaluated Degree-based Hashing (DBH) [15] and High-Degree Replicated First (HDRF) [12] – two of the best-performing strategies regarding partitioning latency and quality [15], [12], [7]. For HDRF, unless stated otherwise, we set the balancing factor $\lambda = 1.1$ as recommended by the authors [12]. We integrated ADWISE as well as DBH and HDRF into the `GrapH` graph processing engine [22] to execute the graph algorithms on the partitioned graphs. Unless stated otherwise, on each of the 8 machines of the compute cluster, each instance of a partitioner (ADWISE, DBH, or HDRF) is loading a disjunct chunk of $1/8$ of the complete graph with a partitioning spread of 4; this makes a total of 32 partitions of the graph.

*A. Efficacy of ADWISE to Minimize Total Graph Latency*

The main idea of ADWISE is to invest more time into graph partitioning to improve the partitioning quality, while reducing the sum of partitioning and processing latency, denoted as the *total graph latency*. In the following, we validate that making the trade-off between partitioning latency and quality controllable—via the partitioning latency preference $L$—yields a reduction of total graph latency by up to 23% compared to HDRF and by up to 47% compared to DBH when computing standard graph algorithms on large real-world graphs.

*1) Brain Graph:* We performed the first set of experiments on the brain graph with moderate clustering coefficient, i.e., there are relatively strong communities in the graph. We executed the PageRank algorithm [4] after partitioning Brain using DBH, HDRF and ADWISE with different (increasing) latency preferences. To evaluate the impact of partitioning quality on graph processing latency, we measured graph processing latency for blocks of 100 iterations of PageRank execution and stacked these blocks on top of the graph partitioning latency to visualize the composition of total graph latency (cf. Figure 7a). We state the measured latency $\bar{L}$ above the respective bars and the latency preference $L$ below the x-axis. This way, the trade-off between partitioning latency and processing latency in ADWISE becomes evident. The most prominent observation is that ADWISE reduces total graph latency by up to 18% compared to HDRF and by up to 39% compared to DBH. Clearly, higher graph processing run-time makes it increasingly beneficial to invest more time into partitioning.

The PageRank algorithm is lightweight in terms of communication and computation: vertices exchange numerical values and perform simple arithmetic calculations in each iteration. To test communication- and computation-heavy graph processing algorithms, we executed an algorithm that solves the NP-complete subgraph isomorphism (SI) problem [11]. We searched Brain consecutively for three subgraphs: circles of different lengths (i.e., path lengths of 19, 15, and 21). Figure 7d visualizes the resulting processing latencies as stacked processing and partitioning latencies as done previously. The sweet spot of minimal total graph latency is clearly visible for ADWISE with $L = 281s$. ADWISE reduces total graph latency by 23% compared to HDRF and by 37% compared to DBH partitioning algorithms. Higher settings of $L$ in ADWISE reduce the graph processing latency of the SI algorithm even further, but do not pay off in terms of total latency.

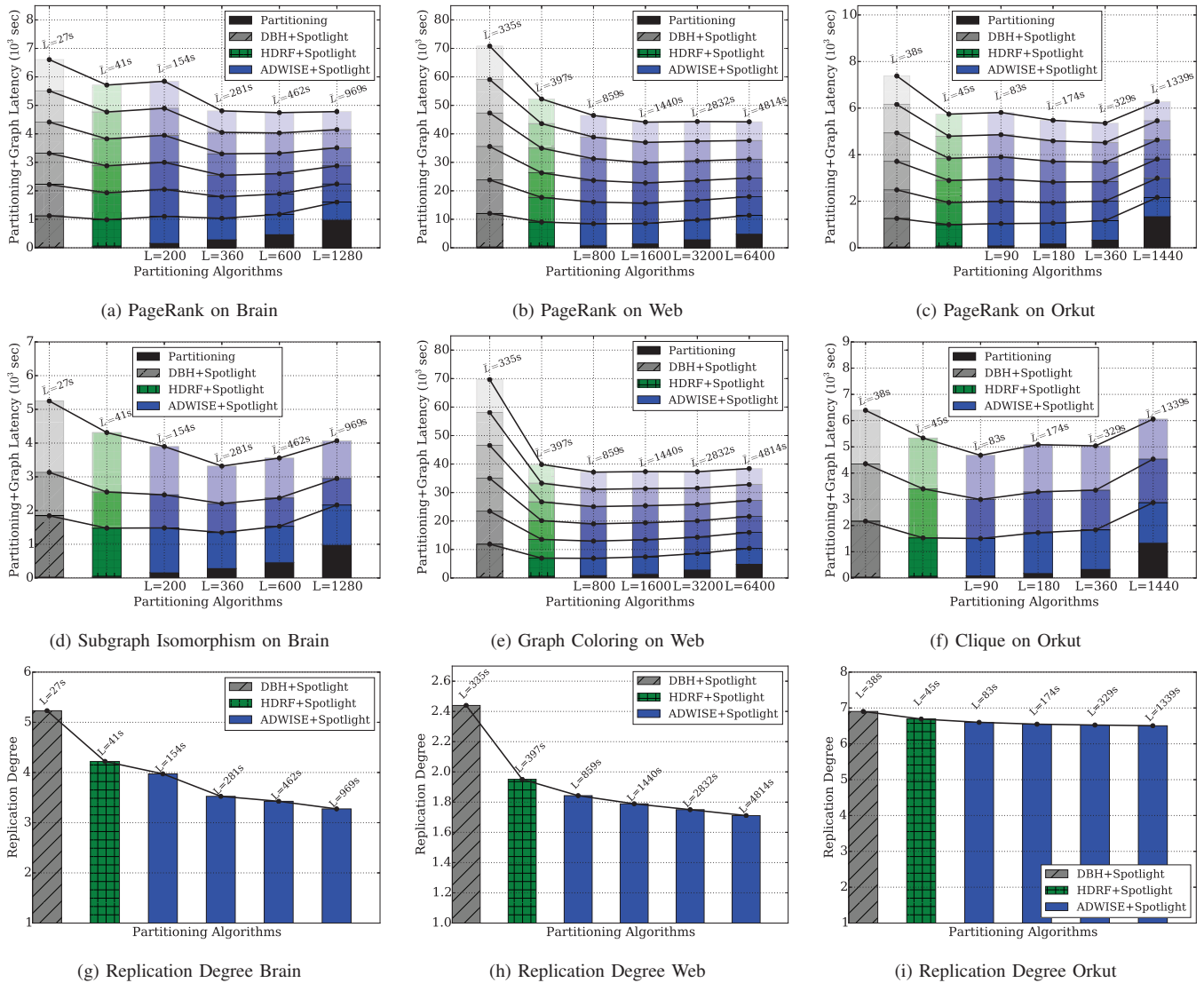

Fig. 7: (a)-(f) Trade-off graph partitioning latency against processing latency. (g)-(i) Replication degree for different partitioning strategies and settings. For all presented results, the partitions are balanced, i.e., $\frac{maxsize-minsize}{maxsize} < 0.05$ (cf. Section III-C).

The reason for reduced graph processing latency when investing more partitioning latency in ADWISE is the improved partitioning quality of the graph. To show this, we plotted the replication degree for the partitioning of the Brain graph in Figure 7g and annotated each experiment with the respective partitioning latency. By increasing the partitioning latency, ADWISE reduces the replication degree subsequently by up to 29% compared to HDRF and by up to 46% compared to DBH. The reduced replication degree leads to reduced communication overhead (i.e., replica synchronization messages) and reduced computational overhead (i.e., replica processing) and therefore directly reduces graph processing latency.

The benefits of reduced graph processing latency outweighed the cost of investing more partitioning latency in the tested real-world workloads on the Brain graph, which experimentally supports our main hypothesis in this paper. To show that this finding generalizes to other types of graphs and other graph processing algorithms, we provide further evaluations in the following.

*2) Web Graph:* The second set of experiments was performed on the Web graph that exhibits a high clustering coef-

ficient. We measure the impact of different latency preferences in ADWISE on the total graph latency in Figure 7b for the PageRank algorithm. ADWISE reduces total graph latency by 16% compared to HDRF and by 38% compared to DBH. Moreover, it is already beneficial to use ADWISE with latency preference $L = 800s$ even for the first 100 iterations. When the graph processing runtime increases (i.e., more iterations are performed), it becomes more and more beneficial to invest more latency into partitioning. Note that as we do not aim for exactly matching the partitioning latency preference, it is possible that we overshoot the preference slightly—in our experiments by up to 7%. In most cases, our adaptive window approach keeps the preference by a wide margin of $20-30\%$.

To test efficacy of ADWISE on other graph processing algorithms, we also executed the graph coloring algorithm presented in [4] (cf. Figure 7e); the graph processing latency was measured after each block of 50 iterations of the graph coloring algorithm. The results show that ADWISE reduces total graph latency at latency preference $L = 800s$ by 9% compared to HDRF and by 47% compared to DBH after 300 iterations of the graph coloring algorithm. Even when executing only a single block of 50 iterations, ADWISE with latency preference $L = 800s$ already reduces total graph latency slightly compared to HDRF and significantly compared to DBH.

The partitioning quality for the different algorithms and settings is depicted in Figure 7h. Investing more partitioning latency in ADWISE reduces replication degree compared to HDRF by 12% (compared to DBH by 41%) for latency $\bar{L} = 859s$ and by 25% (compared to DBH by 51%) for latency $\bar{L} = 4814s$. As expected, allowing for larger partitioning latency in ADWISE leads to larger window sizes which leads to more informed partitioning decisions.

These evaluations on the billion-scale Web graph support our initial hypothesis that the trade-off between partitioning and graph processing latency is not addressed optimally by existing single-edge streaming algorithms. ADWISE proved its efficacy to reduce total graph latency when applied on the Web graph for both the PageRank and the graph coloring workload.

*3) Orkut Graph:* We performed a third set of experiments on the Orkut social network graph. Orkut has a low clustering coefficient, so that the clustering score in ADWISE is not effective and, hence, was switched off for this graph. For the PageRank algorithm, improving partitioning quality with ADWISE leads to decreased total graph latency by up to 11% compared to HDRF and up to 29% compared to DBH (cf. Figure 7c). Clearly, investing more partitioning latency in ADWISE pays off in comparison to single-edge streaming.

For generality, we also evaluated the graph problem of finding fixed-sized cliques in the graph (cf. Figure 7f). We searched for cliques of sizes 3, 4, and 5 with a random walker based clique algorithm: vertices exchange messages of partially found cliques and probabilistically (P = 0.5) forward these messages if they are connected to all vertices in the partial clique message (probabilistic flooding). We repeated the computation ten times for each clique size, starting the random-walk algorithm at ten different randomly chosen vertices. As the results show, ADWISE achieves the minimal total graph latency at partitioning latency $\bar{L} = 83s$ with total latency reduced by 13% compared to HDRF. The larger partitioning latency settings $\bar{L} = 174s$ or $\bar{L} = 329s$ still reduce end-to-end latency slightly compared to HDRF. For even larger partitioning latencies, total graph latency increases due to the more and more prominent impact of the partitioning latency.

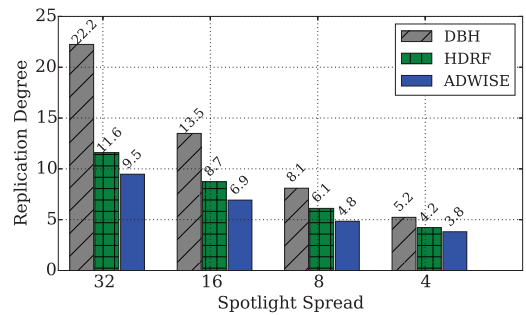

Fig. 8: Efficacy of spotlight optimization on Brain.

In comparison to the other two graphs, Orkut's replication degree is on a relatively high level for all partitioning algorithms (cf. Figure 7i). The reason is that the Orkut graph has a very low clustering coefficient: There is little locality in the edge stream that can be exploited by streaming partitioning algorithms. Still, ADWISE reduces replication degree by up to 4% compared to HDRF and by up to 7% compared to DBH. As shown previously, this small reduction of replication degree leads to significant reductions of graph processing latency. We attribute this effect to the observation that some replicas contribute to overall communication overhead much more than others [11]. Improving locality of a few of those replicas can result in super-linear reductions of graph processing latency.

**Result discussion:** Our experiments on three real-world graphs from different domains using four basic graph processing algorithms validate that single-edge streaming partitioning algorithms are not able to solve the trade-off between partitioning latency and graph processing latency optimally. ADWISE fills this gap by offering the option to invest more time into partitioning to improve the replication degree. This investment pays off in practical use cases – such that the total graph latency can be reduced significantly in our experiments. On the other hand, larger partitioning latencies, e.g. 10 times the single-edge latency, can lead to higher total latency due to the increasing impact of the partitioning latency.

As a practical guideline for users of ADWISE, we propose to invest about three times the latency of single-edge streaming algorithms for graph algorithms with equal or more communication volume as PageRank. If the single-edge streaming latency is not known or can not be estimated, it would even pay off to run a single-edge streaming algorithm once to measure the latency and then invest twice this latency into ADWISE.

*B. Spotlight*

Finally, we experimentally validated efficacy of the spotlight optimization (cf. Figure 8). We measured replication degree using the same computing infrastructure for all three partitioning strategies, i.e., DBH, HDRF, and ADWISE. We varied the spread of the spotlight optimization, i.e., the number of disjoint out-partitions of the $z = 8$ partitioners (cf. Section III-D). Clearly, smaller spread values lead to greatly reduced replication degree by up to $76\%$. The spotlight optimization is extremely effective for *all* initial partitioning strategies: it reduces replication degree significantly. Existing systems [12], [15] use a maximal spread size (e.g. spread of 32 for $k = 32$ partitions) which makes parallel graph loading less effective.

## V. RELATED WORK

The idea of solving large-scale graph problems in a streaming fashion is well-documented in literature [23], [24]. Stanton and Kliot [25] firstly proposed several **edge-cut partitioning** heuristics working in one pass over the graph vertices. FENNEL [26] places the vertex on a partition with many neighbors and few non-neighbors. Nishimura and Ugander [27] proposed a restreaming model that improves the partitioning in each pass using information from the previous pass. METIS [28], an iterative multi-level partitioning algorithm, produces high-quality edge-cuts for graphs with a few million edges [29] but does not scale to massive graphs [26]. Wang et al. [30] proposed a distributed partitioning algorithm based on multi-level label propagation. Zheng et al. [31] consider heterogeneous infrastructures, and Shang et al. [32] heterogeneous workloads. Martella et al. [33] proposed Spinner, a distributed edge-cut partitioning algorithm on top of the Pregel API [3] that migrates vertices to adapt the partitioning dynamically. However, all of these algorithms perform edge-cut partitioning which can not be converted to a good vertex-cut partitioning [8]. For example, the number of edges to be cut in a star-like graph with $|E|$ edges is in $\Omega(|E|)$ – while the number of vertices to be cut is in $O(1)$. ADWISE employs vertex-cut partitioning.

Several streaming **vertex-cut partitioning** algorithms have been proposed. Many graph processing systems use *hashing* [5], [4] which is fast and leads to good workload balancing, but also to high replication degree, graph processing latency and communication overhead. GraphBuilder [34] is a grid-based hashing solution restricting replication of each vertex to a subset of the partitions. Degree-based hashing (DBH) [15] assigns edges to partitions by hashing only the low-degree vertex of an edge leading to better clustering properties. GraphA [35] proposes the use of an incremental number of vertex hash functions to ensure that low-degree vertices are assigned to the same partition and no large imbalances arise. The idea of 1D (and 2D) partitioning [5] is to perform edge assignments based on the adjacency matrix, i.e., assigning all edges based on the row (and column) of their source (and destination) vertex. In contrast to the previous algorithms, Greedy [4] assigns edges to partitions by considering locality explicitly, i.e., where incident vertices are already replicated.

The degree-aware algorithm HDRF [12] (i.e., high-degree vertices are replicated first) is one of the best streaming vertex-cut algorithms outperforming even offline algorithms. PowerLyra [9] extends Greedy to hybrid-cuts by cutting high-degree vertices and edges incident to low-degree vertices. HoVerCut [13] enables multi-threaded processing of the graph stream by granting batch-wise, parallel access to the shared state of the partitioning algorithm. H-load [22] and G-cut [36] consider heterogeneous environments to minimize overall graph processing costs. These vertex-cut algorithms perform a single pass over the edge stream. We have shown that this extreme choice in the search space between low partitioning and low graph processing latency is not optimal for many real-world graph processing tasks. However, as a benchmark we selected the best partitioning algorithm for many graphs, i.e., HDRF, based on an experimental comparison of a wide range of single-edge streaming partitioning algorithms [7]. Note that we did not consider algorithm-specific partitioning strategies using domain knowledge to optimally partition the data for a specific graph algorithm [37]. These methods are not generally applicable to a wide range of graphs and graph applications.

H-move [22] is an iterative communication-aware algorithm that repartitions the graph during graph processing. Rahmanian et al. [38] performed distributed partitioning using an iterative swap heuristic. Huang and Abadi [39] perform dynamic edge-cut partitioning with the possibility of replication, i.e., a hybrid dynamic partitioning algorithm combining edge-cut and vertex-cut. As reassignment of vertices is allowed, the proposed algorithms have super-linear runtime and, hence, do not fit into the streaming edge-cut partitioning model. Zhang et al. [40] developed an interesting all-edge neighborhood expansion (NE) heuristic with polynomial runtime that grows each partition separately using a proximity function. The authors proposed to apply NE iteratively on a random graph sample to reduce memory consumption, but there is no examination of how varying the graph sample size impacts partitioning latency and quality. Studying this trade-off is the main goal in this paper. To summarize, all of these algorithms are computationally more intensive with **super-linear runtime** and therefore not suitable for an initial loading of the graph.

## VI. CONCLUSION

Distributed graph systems rely on fast and effective partitioning algorithms. In recent years, single-edge streaming partitioning dominated the landscape of partitioning algorithms due to the linear runtime complexity. This paper proposes the window-based streaming partitioning algorithm ADWISE that allows for investing more partitioning latency to improve partitioning quality and thus, reduce graph processing latency. ADWISE reduces total end-to-end latency by up to $23 - 47\%$ compared to single-edge streaming in multiple realistic scenarios. Moreover, the novel spotlight optimization — a simple tweak that can be applied to any partitioning algorithm with parallel loading — reduces the replication degree of all evaluated partitioning strategies by $3-4\times$ without introducing computational overhead.